# Ordering Mixed-Q Topological Magnetism into Lattice via Moiré Engineering

Xiudong Wang, Zhonglin He, Kaiying Dou, Ying Dai*, Baibiao Huang, and Yandong Ma*

School of Physics, State Key Laboratory of Crystal Materials, Shandong University, Shandanan Street 27, Jinan 250100, China

*Corresponding author: daiy60@sdu.edu.cn (Y.D.); yandong.ma@sdu.edu.cn (Y.M.)

**Abstract**

Topological magnetic lattices offer a fertile ground for exploring fundamental physics and developing novel spintronic devices. However, current research is predominantly confined to single-Q topologies hosting uniform type of quasiparticle. The realization of exotic mixed-Q states, where distinct topological quasiparticles co-assemble into an ordered lattice, remains largely unexplored. Here, we propose a generic mechanism to order disordered mixed-Q topological magnetism into periodic lattice via moiré engineering. By leveraging the synergy between spatially modulated interlayer coupling and intrinsic intralayer magnetic frustration, we demonstrate that moiré potential can effectively regularize skyrmions, antiskyrmions, and magnetic bubbles into a hybrid lattice. Combining first-principles with atomistic spin simulations, we validate this mechanism in twisted bilayer $CrGaTe_3$, identifying it as an exemplary platform for hosting these complex ordered textures. We systematically map the phase evolution as a function of twist angle and biaxial strain, unveiling the critical role of moiré potential in stabilizing mixed-Q lattice. Our findings significantly advance the frontier of topological and moiré spintronics.

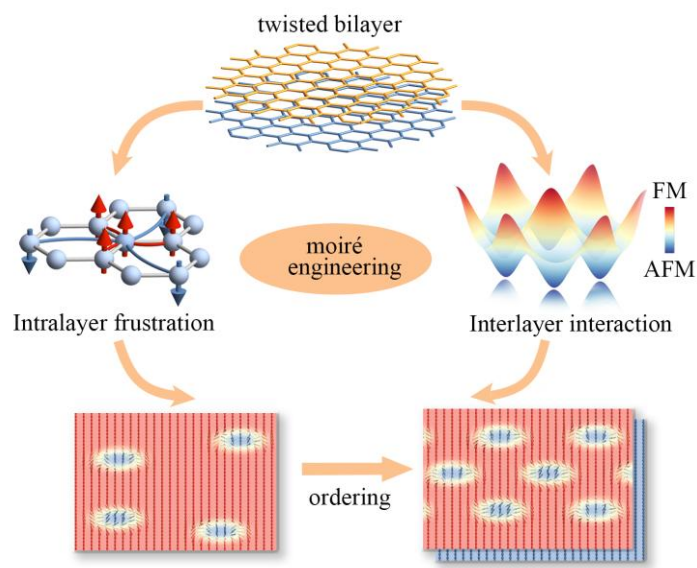

**Introduction**

Topological magnetism, manifested as swirling spin textures with non-trivial topological charge (Q), has garnered intense interest due to its particle-like stability and potential for next-generation spintronics [1-6]. These quasiparticles can self-assemble into ordered topological magnetic lattices (TMLs), providing an ideal platform for exploring their collective behaviors. TMLs facilitate unconventional reservoir computing networks [7,8], skyrmion reshuffling chambers [9,10] and token-based Brownian computing [11]. Beyond applications, they exhibit exotic physical phenomena, including the topological Hall effect [12-14], skyrmion Hall effect [15-17], and multiple phase transitions [18,19]. However, existing research on TMLs is predominantly confined to single-Q topologies hosting a uniform quasiparticle species [20-23]. In contrast, mixed-Q lattices, where distinct spin textures co-assemble, are anticipated to exhibit far richer physical behaviors and broader functionalities [24-26]. Nevertheless, due to stringent stability requirements, realizing such ordered mixed-Q states remain a formidable challenge, leaving this frontier largely underexplored.

Recently, moiré engineering has emerged as a transformative paradigm for uncovering novel physical phenomena. The long-period moiré pattern introduces a periodic modulation of local stacking configurations, imposing a new length and energy scale dictated by the twist angle. This tunability provides an additional degree of freedom for spatially tailoring materials properties, leading to unexpected discoveries such as topological insulator states [27], unconventional superconductivity [28], moiré excitons [29] and polar topology [30]. Beyond electronic characteristics, recent studies demonstrate that moiré superlattices can also effectively engineer topological magnetic properties [31-40]. However, the potential of utilizing moiré potential to regularize and stabilize disordered topological textures into ordered mixed-Q lattices remains an open and critical question.

In this Letter, we bridge this gap by establishing a general framework for ordering mixed-Q topological magnetism via the moiré effect. Our strategy explicitly exploits the interplay between the periodically varying interlayer exchange field—imposed by the twisted interface—and the inherent magnetic frustration of the host material. We reveal that the moiré superlattice can effectively pin and organize skyrmions, antiskyrmions, and magnetic bubbles into an ordered array. We instantiate this mechanism in twisted bilayer $CrGaTe_3$ using a combination of first-principles calculations and large-scale spin dynamics simulations. By systematically mapping the magnetic phase diagram, we elucidate how the twist-angle-dependent moiré potential acts as a tunable virtual field to stabilize diverse topological textures. Our work not only demonstrates the feasibility of mixed-Q TML but also provides a roadmap for designing high-density topological memory architectures.

**Results and Discussion**

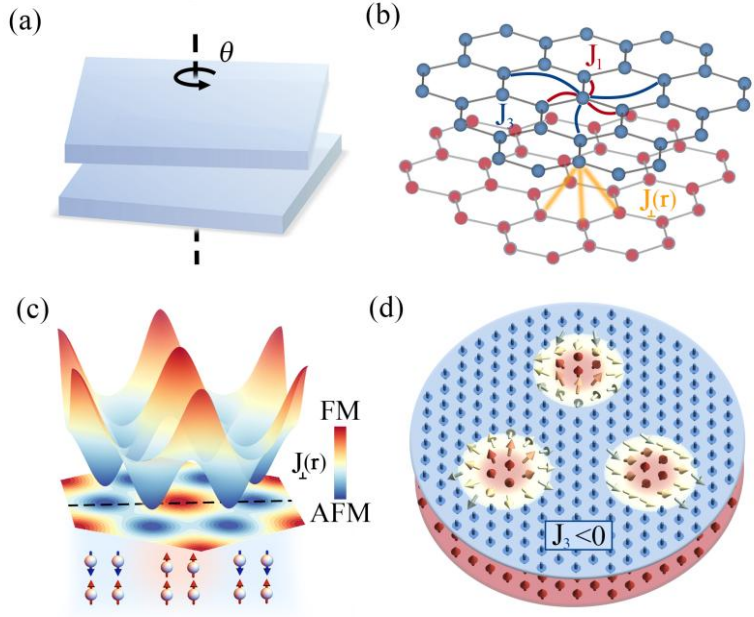

**Fig. 1.** Schematic of ordering mixed-Q topological magnetism into lattice though moiré engineering. (a) Illustration of a twisted van der Waals bilayer with a twist angle $\theta$. (b) Magnetic atomic lattice model considering intralayer exchange interaction ($J_1$-$J_3$ model) and interlayer exchange coupling $J_\perp(r)$. (c) Spatial landscape of the moiré-modulated interlayer coupling $J_\perp(r)$. The color bar indicates FM and AFM regions. The bottom panel illustrates the corresponding interlayer spin alignment along the dashed line. (d) Spin textures of mixed-Q topological magnetic lattice (mixed-Q TML) in twisted bilayer.

To elucidate the mechanism underlying the moiré-engineered ordering of mixed-Q topological magnetism in twisted van der Waals bilayers [**Fig. 1(a)**], we employ a general Heisenberg Hamiltonian to capture the magnetic properties:

$$H = \sum_{t=u,l} H^t_{intra} + H_{inter},$$

where $H_{inter}$ represents interlayer exchange interaction. $H^t_{intra}$ denotes the intralayer exchange interaction for upper ($t=u$) or lower ($t=l$) layer:

$$H^t_{intra} = -J_1\sum_{i,j} \mathbf{S}_i \cdot \mathbf{S}_j - J_2\sum_{i,j} \mathbf{S}_i \cdot \mathbf{S}_j - J_3\sum_{i,j} \mathbf{S}_i \cdot \mathbf{S}_j - \sum_{\langle i,j \rangle} \mathbf{D}_{ij} \cdot (\mathbf{S}_i \times \mathbf{S}_j) - K\sum_i (S_i^z)^2.$$

Here, $\mathbf{S}_{i(j)}$ denotes the normalized spin at the site $i$ ($j$). $J_1$, $J_2$ and $J_3$ represent the nearest-neighbor (NN), next-nearest-neighbor (NNN) and third-nearest-neighbor (3NN) intralayer exchange interactions, respectively. $K$ denotes the single-ion anisotropy and $\mathbf{D}_{ij}$ is the Dzyaloshinskii-Moriya interaction (DMI) vector between

NN $S_i$ and $S_j$.

To characterize the resulting spin textures, we parameterize the normalized magnetization vector field $n(r)$ in spherical coordinates as $n(r) = [sin\,\Theta(r)\,cos(m\phi+\gamma), sin\,\Theta(r)\,sin(m\phi+\gamma), cos\,\Theta(r)]$, where $\Theta(r)$ is the radial polar angle function, while $m$, $\gamma$, $\phi$ denotes vorticity, helicity and spatial azimuthal angle, respectively. We further quantify the topology by the topological charge $Q = \frac{1}{4\pi}\int n(r)\cdot\left(\frac{\partial n}{\partial x}\times\frac{\partial n}{\partial y}\right)d^2r$, which counts the integer winding number $n(r)$ wrapping the unit sphere.

While topological spin textures such as skyrmions often arise from the competition between DMI and ferromagnetic (FM) exchange [41-44], DMI typically locks the spin chirality, thereby hindering the diversity of topological magnetism required for mixed-Q state. Consequently, we focus on centrosymmetric systems (where $D_{ij} = 0$) described by a frustrated $J_1$-$J_3$ model [**Fig. 1(b)**]. In this framework, with $J_1 > 0$ favoring FM alignment, the sign of $J_3$ becomes pivotal. For $J_3 > 0$, all intralayer interactions cooperatively stabilize a collinear FM ground state. Conversely, for $J_3 < 0$, the antiferromagnetic (AFM) nature of long-range interaction competes with the FM $J_1$. This induced magnetic frustration liberates the spin textures from chirality constraints, stabilizing diverse nontrivial quasiparticles such as skyrmions (Q=$\pm 1$, $m=1$), antiskyrmions (Q=$\pm 1$, $m=-1$), and magnetic bubbles (Q=0, $m=0$) [26,42,45-47].

Upon stacking two such monolayers to form a twisted bilayer, the interlayer exchange interaction becomes a curial determinant of the resulting spin textures. The interlayer exchange term is given by:

$$H_{inter} = -\sum_{i,j} J_\perp(r_{ij}) S_i^u \cdot S_j^l,$$

where $J_\perp(r_{ij}) = J_\perp(r_i^u - r_j^l)$ denotes the position-dependent interlayer coupling between spins $S_i^u$ and $S_j^l$. The formation of a moiré superlattice creates a periodic variation in the local stacking registry, which imposes a strong spatial modulation on the interlayer coupling $J_\perp(r_{ij})$. Crucially, $J_\perp(r_{ij})$ inherits the periodicity of the moiré superlattice, i.e., $J_\perp(r_{ij}) = J_\perp(r_{ij}+R)$, where $R$ is the lattice constant of moiré superlattice.

This modulation typically manifests in two distinct scenarios: (i) a magnitude-modulated regime, where $J_\perp(r_{ij})$ maintains a constant sign [either FM or AFM] across the entire supercell; or (ii) a sign-alternating regime, where $J_\perp(r_{ij})$ oscillates between FM [$J_\perp(r_{ij}) > 0$] and AFM [$J_\perp(r_{ij}) < 0$] character depending on the local registry, as depicted in **Fig. 1(c)**. We anticipate that the latter scenario characterized by an FM-AFM alternating $J_\perp(r_{ij})$ generates a significantly deeper moiré potential landscape. In this context, supposing the AFM-coupled regions are energetically preferred and spatially dominant, the higher-energy FM-coupled regions would be forced into geometrically isolated domains. These FM islands would serve as localized potential wells that effectively trap and organize the disordered topological spin textures. Consequently, the moiré superlattice can serve as a template to pin and order these diverse quasiparticles into a lattice, i.e., mixed-Q topological magnetism lattice (mixed-Q TML) [**Fig. 1(d)**].

Guided by this theoretical framework, the material realization of mixed-Q TMLs imposes three critical constraints on the candidate system: (i) the constituent monolayer must possess intrinsic magnetic frustration (specifically, a considerable AFM $J_3$) to support diverse topological excitations; (ii) the twisted bilayer must exhibit a sign-alternating interlayer coupling [$J_\perp(r_{ij})$] to create the deep moiré potential and isolated domains required for lattice stabilization; and (iii) the AFM-coupled regions are energetically preferred and spatially dominant. Identifying a candidate that simultaneously satisfies these stringent conditions is non-trivial. Through a systematic search of two-dimensional van der Waals magnets, we identify CrGaTe$_3$ as a candidate system.

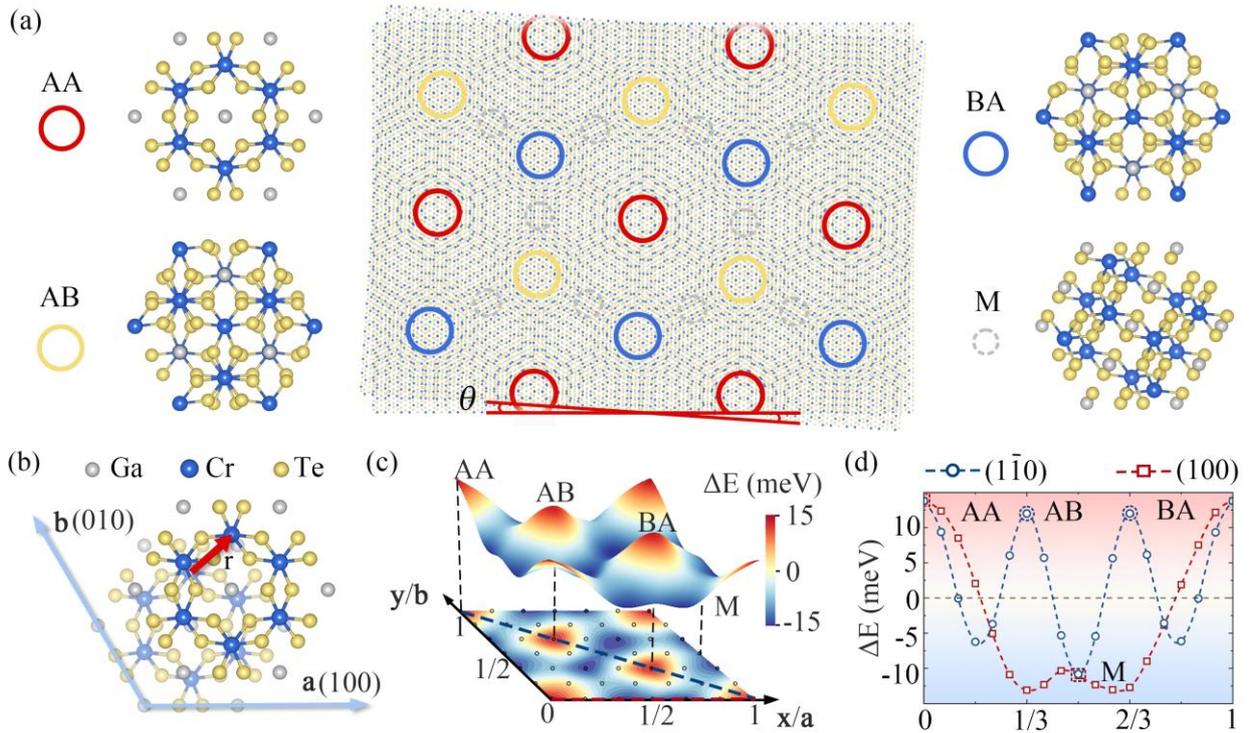

**Fig. 2.** Stacking-dependent interlayer coupling in twisted bilayer CrGaTe$_3$. (a) Top view of the crystal structure at twist angle $\theta$. Surrounding panels depict four high-symmetry local stacking configurations: AA (red), AB (yellow), BA (blue) and M (grey). (b) Definition of the interlayer translation vector $r$, which characterizes the local atomic registry within the moiré supercell. (c) Three-dimensional landscape and projected contour map

of the interlayer coupling energy $\Delta E(r_i)$. (d) Line profiles of $\Delta E(r_i)$ as a function of interlayer translation along [100] (red-dashed line with squares) and [1$\bar{1}$0] (blue-dashed line with dots) directions.

**Fig. S1** presents the crystal structure of monolayer (ML) CrGaTe$_3$, which crystallizes in a hexagonal lattice with the space group P$\bar{3}$m1(D$_{3d}$). The structure features a central Cr atom layer sandwiched between the two Ga/Te atomic layers. Each Cr atom is surrounded by six Te atoms, forming a slightly distorted octahedron. The Cr sublattice arranges into a honeycomb network with additional Ga-Ga dimers located vertically at the hollow sites of the Cr hexagons. The optimized lattice constant is 6.97 Å, fitting well with previous work [48]. The stability of monolayer CrGaTe$_3$ is assessed by calculating its phono spectrum. As shown in **Fig. S2**, the absence of imaginary modes across the entire Brillouin zone confirms its dynamic stability.

To satisfy charge neutrality, the Cr atoms adopt an oxidation state with a $3d^3$ electronic configuration. Under the octahedral crystal field, the $d$ orbitals split into two groups ($e_g$ and $t_{2g}$). The three electrons of Cr atom half-fill the $t_{2g}$ orbital, yielding a magnetic moment of $3\mu_B$ per Cr atom, which is further verified by our calculations. Then we calculate the spin-polarized band structure of ML CrGaTe$_3$. As shown in **Fig. S3**, the band structure exhibits semiconducting nature with an extra narrow indirect band gap of 73 meV.

To investigate the magnetic interactions in ML CrGaTe$_3$, we employ the monolayer Heisenberg spin Hamiltonian $H_{mon}$. Considering the energy difference between various magnetic configurations as depicted in **Note S1**, we extract the exchange parameters: $J_1$=18.51meV, $J_2$=-0.32meV, $J_3$=-4.50meV. These values indicate that the system is well-described by the $J_1$-$J_3$ frustration model, as $J_2$ is negligible. Based on symmetry analysis (Moriya's rule [49]), DMI is forbidden due to the inversion symmetry protection. The single-ion anisotropy $K$ is calculated to be 1.56meV, implying that ML CrGaTe$_3$ possesses strong out-plane magnetic anisotropy. The spin textures of ML CrGaTe$_3$ shown in **Fig. S6** demonstrate the existence of disordered topological quasiparticles.

Having established the properties of ML CrGaTe$_3$, we then turn to the twisted bilayer (TBL) GrGaTe$_3$. In the long-period moiré superlattice, the local stacking configurations can be described by translating the upper layer of the untwisted bilayer by a vector $r = \eta a + \nu b$ [**Fig. 2(b)**], where $\eta, \nu \in [0,1]$, and $a$ and $b$ are the unit-cell lattice vectors. As identified in **Fig. 2(a)**, the local regions with AA, AB, BA and M stacking configurations correspond to the upper layer laterally shifted by $r$ = {0, 2/3$a$+1/3$b$, 1/3$a$+2/3$b$, 1/2$a$($b$)}, respectively. To quantify the interlayer interaction, we adopt $\Delta E(r_i) = (E_{afm}(r_i) - E_{fm}(r_i))/2$ to describe the energy difference between FM and AFM interlayer spin configurations for each stacking pattern [**Fig. 2(c)**]. The interlayer ground state oscillates between FM and AFM as the stacking pattern varies. Specifically, the AA, AB and BA regions favor FM coupling, whereas the M region stabilizes AFM coupling. **Fig. 2(d)** details the variation of $\Delta E$ along the [100] (red dashed line) and [1$\bar{1}$0] (blue dashed line) directions. Notably, along the [100] direction, the coupling switches between AFM and FM only once. In contrast, along the [1$\bar{1}$0] direction, it undergoes three reversals between FM and AFM, indicating a more complex spatial modulation. More importantly, our total energy analysis (**Fig. S7**) reveals the AFM-coupled regions represent the global energy minima of the superlattice and form the dominant background.

Following the methodology established in previous work [34,38,50], we derive the spatially modulated interlayer exchange interactions $J_\perp(r_{ij})$ from the stacking-dependent coupling energy $\Delta E(r_i^{u(l)})$ for the upper (lower) layer. We consider a spin $S_i^{u(l)}$ at position $r_i^{u(l)}$ in the upper (lower) layer interacting with a set of neighboring spins $S_j^{l(u)}$ at positions $r_j^{l(u)}$ in the lower (upper) layer. The relationship between $J_\perp(r_{ij})$ and $\Delta E(r_i^{u(l)})$ can be written as $\sum_j J_\perp(r_{ij}) = \Delta E(r_i^{u(l)})/2$. Assuming $J_\perp(r_{ij})$ as an exponential decaying function of distance, $J_\perp(r_{ij})$ can be written as:

$$J_\perp(r_{ij}) = \Delta E(r_i^{u(l)}) e^{-\delta_i^{u(l)} \sqrt{|r_i^u - r_j^l| + d^2}} / 2,$$

where $d$ denotes the vertical separation between magnetic layers and decay factor $\delta_i^{u(l)}$ is determined by the equation $\sum_j e^{-\delta_i^{u(l)} \sqrt{|r_i^u - r_j^l| + d^2}} = 1$. To adequately capture the spatial distribution of the interaction, we select a cutoff radius $|r_i^u - r_j^l| \leq a$ [50], as exponential terms decays rapidly with distance.

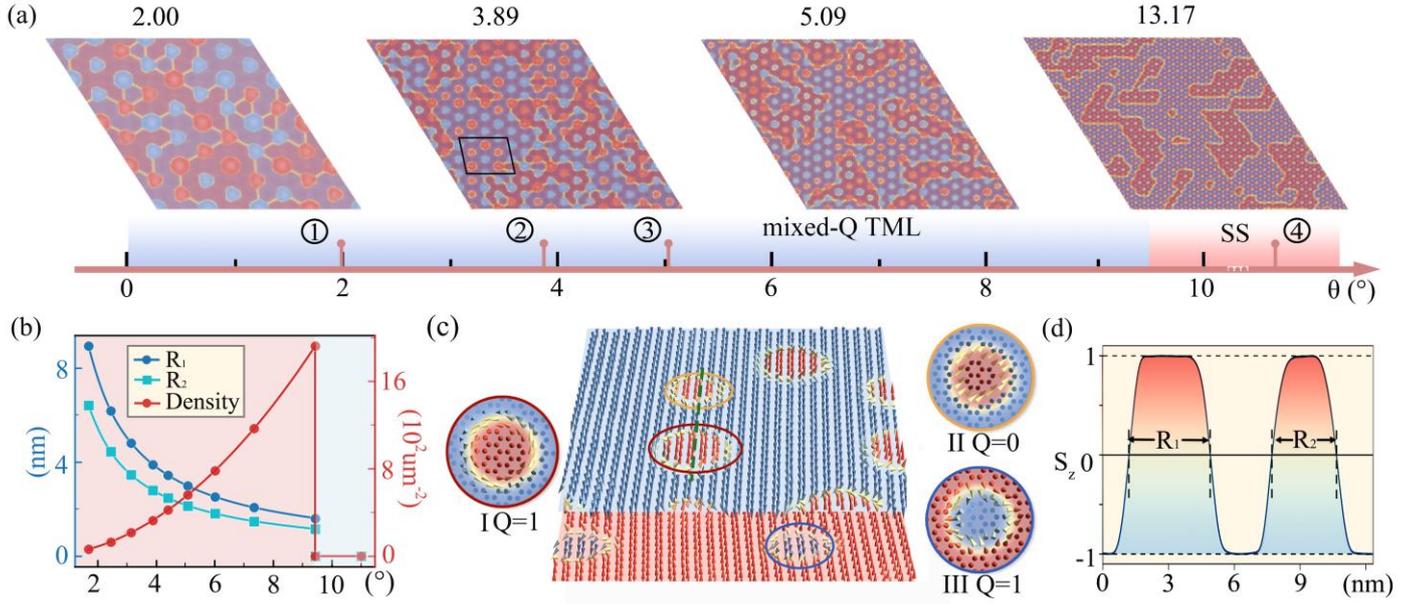

**Fig. 3.** Twist-angle-dependent magnetic phase diagram. (a) Magnetic phase diagram as a function of twist angle $\theta$, identifying two distinct phases: mixed-Q TML and half-bubble lattice (HBL). (b) Quantitative evolution of the quasiparticle radii ($R_1$, $R_2$) and their packing density as a function of twist angle. (c) Detailed spin textures of mixed-Q TML state at $\theta = 3.89°$, corresponding to the region marked by black rectangular outline in (a). The enlarged view highlights the coexistence of three distinct topological quasiparticles: skyrmion (I, red circle), bubble (II, orange circle) and antiskyrmion (III, blue circle). (d) Cross-sectional profiles of the out-of-plane spin component ($S_z$) along the green dashed lines in (b).

By incorporating the extracted magnetic parameters into the Hamiltonian, we perform atomistic spin model simulations based on the Landau-Lifshitz-Gilbert (LLG) equation to systematically investigate the spin textures of twisted bilayer CrGaTe$_3$ as a function of twist angle $\theta$. The twist angle $\theta$ is intrinsically linked to the moiré superlattice vector $\boldsymbol{R}$. To preserve the translational symmetry of the moiré superlattice, $\boldsymbol{R}$ must satisfy the commensurability condition $\boldsymbol{R}=k\boldsymbol{a}+h\boldsymbol{b}$, where $k, h \in \mathbb{N}$. Consequently, the moiré period length $|\boldsymbol{R}|$ and the twist angle $\theta$ (**Note S2**) are derived as functions of $k$ and $h$:

$$|\boldsymbol{R}| = |k\boldsymbol{a}+h\boldsymbol{b}| = a_0\sqrt{k^2+h^2-kh},$$

$$\theta = \frac{180°}{\pi}\arccos\left[1-\frac{3h^2}{2(k^2+h^2-kh)}\right].$$

To simplify the parameter space, we fix $h=1$ and vary the integer $k$ to systematically tune $\theta$.

As illustrated in **Fig. 3(a)**, the magnetic ground state evolves distinctly across two regimes. In the small twist angle regime (0°<$\theta$<9.37°), the moiré superlattice period is large, resulting in a slowly varying interlayer coupling $J_\perp(r_{ij})$. This spatial modulation, in synergy with intralayer $J_1$-$J_3$ interactions, stabilizes an ordered mixed-Q TML. By correlating the spatial distribution of quasiparticles with local moiré stacking order (**Fig. S7**), it is evident that they preferentially nucleate within the AA, AB and BA regions. In these domains, the interlayer interaction is FM, creating effective potential traps for topological excitations.

Notably, the geometric extent of these FM-coupled interlayer region varies with the stacking patterns: AA stacking regions are spatially larger than the AB and BA regions. Consequently, the nucleated quasiparticles exhibit distinct sizes, with the radius $R_1$ in AA regions larger than $R_2$ in AB and BA regions. As the twist angle increases, the moiré periodicity $|R|$ shrinks, imposing a stronger confinement on the spin textures. This leads to a systematic scaling of the mixed-Q TML: the radii $R_1$ ($R_2$) decreases inversely from 8.93 (6.41) nm to 1.61 (1.15) nm, while the packing density exhibits a quadratic increase from 0.62 to $19.23 \times 10^2$ um$^{-2}$ as twist angle $\theta$ scales from 1.70° to 9.37° [**Fig. 3(b)**].

A critical transition occurs when the twist angle exceeds 9.37°. At this threshold, the spatial extent of the FM-coupled AA, AB and BA regions shrinks below the characteristic magnetic length scale required to sustain finite-sized quasiparticles. This geometric confinement triggers a sudden collapse of the topological state, where both radii and density drop abruptly to zero. Consequently, the topological excitations are suppressed, and the system reverts to a half-bubble lattice (HBL) state with interlayer AFM alignment.

To elucidate the microscopic nature of the mixed-Q TML phase, we focus on the representative state at $\theta$=3.89°. **Fig. 3(c)** presents an enlarged view of the spin texture within the selected region. We observe that the topological quasiparticles manifest in either the upper or lower layer with equal probability and their core polarization exhibit a stochastic up-or-down orientation owing to the energetic degeneracy between the two layers. The mixed-Q TML is composed of three distinct species categorized by their topological winding properties. In Region I, the out-plane spin components at the core are oriented upward, while the in-plane spin components wind continuously around the core, co-rotating with the spatial azimuthal angle. This forms a vortex-like spin texture typically identified as a skyrmion (Q=1, m=1). In Region II, the spin texture lack topological winding. Despite the reversal of the out-of-plane at the core, the in-plane components retain a uniform orientation, characterizing a topologically trivial magnetic bubble (Q=0, m=0). In Region III, the out-plane spin components at the core are reversed relative to region I and the in-plane spin components exhibit a counter-rotating behavior with respect to the spatial azimuthal angle. This results in a saddle-like spin texture identified as an antiskyrmion (Q=1, m=-1). Crucially, each species can be stabilized in either AA region or AB/BA region with different scales. As evidenced in **Fig. 3(d)**, two characteristic radii are observed: a larger radius of $R_1$=7.7 nm in the AA region, and a smaller radius of $R_2$=5.6 nm in the AB and BA regions. Given this stable coexistence of quasiparticles with distinct topological charges and varying scales, the mixed-Q TML is firmly confirmed in twisted bilayer CrGaTe$_3$.

We further explore the influence of external magnetic field $B_z$ on the evolution of spin textures. As depicted in **Fig.S10**, as $B_z$ increases, the AFM-coupled regions remain stable, while the spin polarization in the FM regions increasingly aligns with direction of the applied field. The population of quasiparticles with downward cores decreases as the magnetic field breaks the energetic degeneracy between the upward and downward configurations. Our observations indicate that at a smaller twist angle ($\theta$=2.00°), the system is more vulnerable to the external magnetic field, as it transitions into a pure mixed-Q TML at 1T. In contrast, at larger twist angle ($\theta$=3.89° and 5.09°), a higher field of 3T is required to orient all quasiparticles upward.

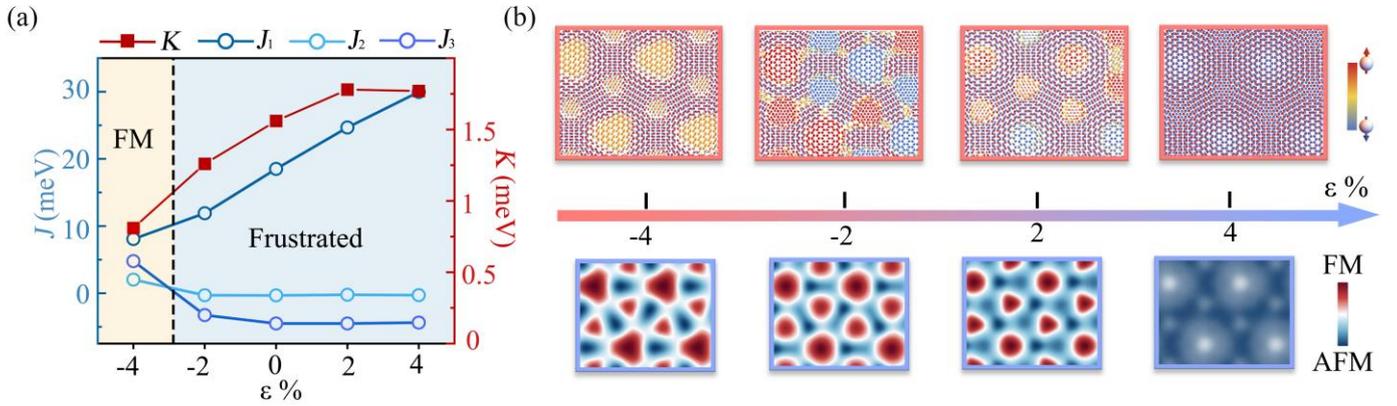

**Fig. 4.** Strain-engineered magnetic interactions and spin textures evolution in twisted bilayer CrGaTe$_3$ at twist angle $\theta$=3.89°. (a) Strain dependence of magnetic parameters in monolayer CrGaTe$_3$: nearest-neighbor ($J_1$), next-nearest-neighbor ($J_2$), third-nearest-neighbor ($J_3$) exchange interaction and single-ion anisotropy ($K$). (b) Bottom panel: Interlayer coupling energy $\Delta E(r_i)$ as a function of applied biaxial strain. Top panel: Evolution of spin textures in the twisted bilayer under biaxial strain from -4% to 4%.

To explore the tunability of these topological states, we map the evolution of spin textures with biaxial strain at a representative twist angle $\theta$=3.89°. As shown in **Fig. 4(a)** and **Table. S1**, $J_1$ increases monotonically from 8.07 to 29.97 meV as the biaxial strain shifts from -4% to 4%. According to the Goodenough-Kanamori-Anderson rules [51], this trend can be attributed to the intricate competition between FM superexchange and AFM direct exchange. The nearly orthogonal Cr-Te-Cr angles (**Table. S2)** ensure the FM superexchange remains the dominant component of $J_1$. As the biaxial strain increases, the expansion of the Cr-Cr distance reduces the direct orbital overlap, thereby suppressing the AFM direct exchange. As the AFM direct exchange diminishes more rapidly than the FM superexchange, $J_1$ is effectively enhanced. Simultaneously, single-ion anisotropy $K$ also increases with the rise of biaxial strain. In contrast, the $J_2$ and $J_3$ undergo a sign reversal from positive to negative as the strain exceeds -4%. This transition indicates that while the system maintains strong local FM coupling via $J_1$, the long-range interactions transition into an AFM coupling regime. The resulting competition between the strengthened $J_1$ and the AFM $J_2/J_3$ introduces magnetic frustration, which is essential for the stabilization of complex topological spin textures.

Subsequently, we consider the influence of biaxial strain on the interlayer interaction. As illustrated in bottom panel of **Fig. 4(b)**, the spatial distribution of interlayer coupling energy $\Delta E(r_i)$ is highly sensitive to strain engineering. At moderate strain (between -2% and 2%), the coupling landscape remains qualitatively similar to the unstrained state, with AA and AB/BA stacking regions FM coupled while other regions remain AFM coupled. Combined with the relatively stable single-ion anisotropy $K$ and frustrated Heisenberg exchange interaction, the mixed-Q TML is maintained, demonstrating its robustness against strain. However, when a large tensile strain of 4% is applied, the interlayer coupling energy becomes uniformly negative across all the stacking patterns, indicating a global transition to AFM interlayer coupling. Without the alternating FM-AFM interlayer potential, the mixed-Q TML is suppressed and the system reverts to a common AFM bilayer. Conversely, under a compressive strain of -4%, while the alternating FM-AFM interlayer coupling is retained, the intralayer interactions shift from frustrated regime to purely FM state. This prevents the spins from undergoing a complete reversal. Instead, within the FM-coupled regions, the spins rotate by only 90°, resulting in the formation of HBL rather than mixed-Q TML.

**Conclusion**

In summary, we establish a universal mechanism to order disordered mixed-Q spin textures into periodic lattice by exploiting the synergy between moiré potential and intrinsic frustration in twisted van der Waals bilayers. First-principles calculations and atomistic simulations further demonstrate CrGaTe$_3$ as a promising platform to host mixed-Q TML. These insights significantly enrich the frontiers of moiré magnetism and topological spintronics.

**Methods**

Our first-principles calculations are performed based on density functional theory (DFT) using the projector augmented wave method as implemented in the Vienna Ab initio Simulation Package (VASP) [52-54]. The exchange-correlation interaction is treated by the Perdew-Burke-Ernzerhof (PBE) functional within the generalized gradient approximation (GGA) [55]. Considering the strong correlation effects of transition metal atoms, we adopt the GGA+U method [56] for the $3d$ electrons of Cr atoms. According to previous works [48,57], the effective onsite coulomb interaction parameter U and the exchange interaction parameter J are set to be 5 eV and 1 eV, respectively. A $9 \times 9 \times 1$ k-point mesh is used to sample the Brillouin zone. The cutoff energy is set to 500 eV. Structures are fully relaxed with the convergence criteria of 0.01 eVÅ$^{-1}$ and $1 \times 10^{-6}$ eV for force and energy, respectively. For exchange interaction parameter calculations, $5\times5\times1$ k-point mesh is adopted in a $2\times2$ supercell. The vacuum space is set to 20 Å to avoid the interaction between adjacent layers. The phonon dispersion is calculated using the PHONOPY code [58]. The zero-damping DFT-D3 method of Grimme is utilized for treating the vdW interaction.

Atomistic spin model simulations are performed using the VAMPIRE package based on the spin Hamiltonian and Landau-Lifshitz-Gilbert (LLG) equation [59,60]:

$$\frac{\partial S_i}{\partial t} = -\frac{\gamma}{(1+\lambda^2)}[S_i \times B_{eff}^i + \lambda S_i \times (S_i \times B_{eff}^i)]$$

Here $S_i$ is a unit vector representing the direction of the magnetic spin moment of site $i$, $\gamma$ is the gyromagnetic ratio and $\lambda$ is the damping constant. $B_{eff}^i = -\frac{1}{u_s}\frac{\partial H}{\partial S_i}$ is the net magnetic field on each spin. To obtain the equilibrium spin textures, the system is gradually cooled from an initial disordered state at 600 K to 0 K. The field-cooling process employs a Gaussian cooling function with a 100 ps cooling time. An ultrafine time step of 0.1 fs is maintained throughout the simulation to ensure the numerical convergence and stability of the complex spin configurations. For the initial state, $10^5$ equilibration steps are performed at 600K to establish a well-defined starting point for the cooling protocol. In simulations, the $\gamma$ and $\lambda$ are set to $1.76 \times 10^{11}$ $T^{-1}s^{-1}$ and 1.0, respectively.

**Supporting Information**

The supporting information is available in **

**Conflict of Interest**

The authors declare no competing financial interest.

**Acknowledgements**

This work is supported by the National Natural Science Foundation of China (No. 12274261) and Taishan Young Scholar Program of Shandong Province.

**Keywords**

topological spin textures, twisted bilayer $CrGaTe_3$, moiré magnetism, mixed-Q topological magnetic lattice, first-principles